# Direct magnetic field detection in the innermost regions of an accretion disc

Jean-François Donati[1], Fréderic Paletou[1], Jérome Bouvier[2] & Jonathan Ferreira[2]

[1]*Lab. d'Astrophysique, Obs. Midi-Pyrénées, F-31400 Toulouse, France*

[2]*Lab. d'Astrophysique de Grenoble, Univ. J. Fourier, F-38041 Grenoble, France*

**Models[1–5] predict that magnetic fields play a crucial role in the physics of astrophysical accretion disks and their associated winds and jets[6,7]. For example, the rotation of the disk twists around the rotation axis the initially vertical magnetic field, which responds by slowing down the plasma in the disk and by causing it to fall towards the central star. The magnetic energy flux produced in this process points away from the disk, pushing the surface plasma outwards, leading to a wind from the disk and sometimes a collimated jet. But these predictions have hitherto not been supported by observations. Here we report the direct detection of the magnetic field in the core of the protostellar accretion disk FU Orionis[8]. The surface field reaches strengths of about 1 kG close to the centre of the disk, and it includes a significant azimuthal component, in good agreement with recent models[5]. But we find that the field is very filamentary and slows down the disk plasma much more than models predict, which may explain why FU Ori fails to collimate its wind into a jet.**

Although magnetic fields have been reported in the external regions of a few protostellar disks[9,10], no field estimate yet exists for the innermost and densest parts of accretion disks from which most of the ejected plasma presumably originates. Obtaining constraints on the field strength and topology in accretion disk cores (such as those derived for cool star surfaces from time-resolved spectropolarimetry[11,12]) would thus be extremely helpful for validating existing models of magnetized accretion/ejection structures around protostars and black holes. Thanks to its high mass-accretion rate[8,13] of about $10^{-4}$ solar masses per year ($M_\odot$ yr$^{-1}$), FU Ori appears as a bright disk completely outshining the central protostar and is thus an obvious candidate for detecting magnetic fields in disk cores. Because models predict typical disk fields of equipartition strength (with roughly equal thermal and magnetic pressures), FU Ori should host fields of several hundred Gauss in its innermost regions, within reach of modern spectropolarimetric techniques[11,14].

During the engineering tests of the new high-efficiency high-resolution spectropolarimeter ESPaDOnS[14], we secured six observations of FU Ori at three different epochs, each consisting of one unpolarized and one circularly polarized spectrum. Using least-squares deconvolution[11] (LSD), we extracted the unpolarized (Stokes *I*) and circularly polarized (Stokes *V*) information from 4,700 spectral lines simultaneously, and produced mean LSD Stokes *I* and *V* profiles of FU Ori at each epoch. Whereas the unpolarized LSD profiles give information about the velocity field of the central regions of FU Ori, the circularly polarized LSD profiles give access to the putative disk magnetic field through the Zeeman signatures it generates in spectral lines. Averaging the four LSD Stokes *V* profiles recorded at the same epoch (30 November 2004), we obtain a clear Zeeman detection in the absorption lines of FU Ori (see Fig. 1); the corresponding line-of-sight magnetic field estimated from the first-order moment of the Zeeman signature[11] is equal to 32±8 G.

This is evidence that the field we detect comes from the disk and not from either the central star or the two recently detected cool companions, each ~100 fainter in the V band than the accretion disk[13,15,16]; it would otherwise require the star to exhibit a 3 kG surface-averaged line-of-sight field and thus to host a large-scale magnetic dipole of polar strength >9 kG (ref. 17), far larger than the global dipolar field component observed on both protostars[18] and young low-mass stars[12]. The LSD Stokes *V* profiles collected at the two other epochs (24 September 2004 and 28 November 2004) are compatible within noise level with the Zeeman signature detected on 30 November (see Fig. 1). As our observing epochs are randomly phased with respect to each other, it suggests that, at first order, the temporal fluctuations from either rotational modulation or intrinsic variability of the disk Zeeman signature are smaller than the signature itself—that is, that the parent disk magnetic structure is grossly axisymmetric.

By applying LSD with line lists corresponding to different spectral types and line strengths (see Fig. 2), we find that the derived LSD profile is closest to a conventional double-peak disk profile when using the weakest features of the G0 line list only. Although symmetric, the derived profile features a conspicuous flat-bottom shape that standard models do not predict. LSD Stokes *I* profiles produced with other line lists are all asymmetric, each at a different level; in addition to the disk profile, they include a spectral contribution strongest at spectral type A0 and blue-shifted with respect to the disk radial velocity of 27 km s$^{-1}$ (refs 19, 20), which presumably forms in a ~10,000 K dense plasma ejected from the disk at speeds up to 80 km s$^{-1}$. This signature is probably due to the strong disk wind of FU Ori[21]. However, the detected Zeeman signature exhibits no such change with the selected line list (and in particular, no blue shift) that correlates with the spectral contribution from the wind (see Fig. 2); we thus conclude that the parent magnetic field is located within the disk only and not in the wind.

We use the LSD Stokes *I* profile produced with the partial G0 list (showing no wind contribution) and the LSD Stokes *V* signature obtained from the full G0 list (containing less noise) to investigate the velocity and magnetic fields of the disk. To this end, we constructed a simple axisymmetric disk model inspired by recent models[5] of magnetic accretion/ejection structures. From the observation that LSD Stokes *V* profiles are two to three times narrower than their unpolarized equivalents (see Figs 1 and 3), we infer that the plasma hosting the parent magnetic field rotates at largely subkeplerian velocities. Outer disk regions rotating at keplerian velocities slow enough (~30–40 km s$^{-1}$) to produce the detected Zeeman signature are indeed too cool (<2,500 K)[13] to contribute more than 1% to the bulk of optical lines[22]. Moreover, they would generate Zeeman signatures much stronger in the red (where cool regions relatively contribute more) than in the blue, in contradiction with observations (see Supplementary Fig. 1). Assuming that a small fraction of the disk plasma rotates at subkeplerian velocities also ensures that the unpolarized profile from the disk is partly filled-in and no longer features a conventional double-peak shape, but rather exhibits a flat-bottom aspect as observed.

The significant velocity shift with respect to line centre of the observed LSD Stokes $V$ profile (~13 km s$^{-1}$) also indicates that the parent disk field includes a significant azimuthal component (see Fig. 3). Both vertical and azimuthal components of the field are inferred by adjusting respectively the antisymmetric and symmetric components of the detected Zeeman signature. A reasonable match to both the unpolarized profile and the circularly polarized profile from the disk is obtained when assuming that the magnetic field occupies ~20% of the disk plasma and that the magnetic plasma rotates two to three times slower than the local keplerian velocity; the vertical component of the derived magnetic field (~1 kG at 0.05 AU) points towards the observer, while the azimuthal component (about half as strong) points in a direction opposite to the orbital rotation.

The field configuration at the surface of FU Ori resembles those predicted by theoretical models of magnetic accretion/ejection structures[3,5,23]. We observe a steady-state magnetic structure persisting in the disk on timescales longer than the dynamical timescale. The detected magnetic field is of equipartition strength, and includes an azimuthal component whose direction agrees with that produced by the vertical field (presumably aligned with the large-scale field) being sheared by the disk orbital motion. The derived azimuthal to vertical field ratio at the disk surface, strongly constrained by the need to match both the symmetric and antisymmetric components of the detected Zeeman signature, is ~0.5 and compatible with expectations (see Fig. 7 of ref. 23), indicating that the large-scale topology is mostly dipolar rather than quadrupolar. We therefore conclude that our observations provide definite evidence, not only that accretion disks are magnetized, but also that the field configuration agrees with those devised by most recent disk/jet models. A poorer match is obtained with disk dynamo models, which predict dominantly toroidal field configurations in inner disk regions[24,25].

However, some differences with model predictions are observed. First, the magnetic field of FU Ori appears filamentary rather than homogeneous, and probably consists of magnetic flux tubes immersed within the bulk of the non-magnetic plasma; although plasma is at rough equipartition inside the tubes, it is essentially unmagnetized in the remaining 80% of the disk, implying an average disk magnetic energy lower than 20% of the disk thermal energy. Second, the observed magnetic plasma is slowed down much more than models expect (see Fig. 5 of ref. 23). It may indicate that the relative disk thickness $h/r$, with which plasma deceleration scales up ($h$ being the disc thickness at radial distance $r$ from disk centre), is significantly larger than the standard value of 0.1, at least in the disk core. This unexpected field property may be a hint as to why FU Ori fails to collimate its wind into a jet (as opposed to other similar objects with detected jets[8]), despite its magnetic topology being apparently favourable for jet-launching.

Our results also demonstrate that accretion disks are successful at triggering turbulent instabilities that produce enhanced radial accretion and drifts of ionized plasma through transverse field lines, as FU Ori would otherwise fail to generate a steady-state axisymmetric magnetic structure similar to what we observe. Although the nature of this instability is still enigmatic[26], our observations suggest that magnetic fields play a significant role in this process, making the magneto-rotational instability[27] and other related instability types[28] likely candidates.

## Methods

### Spectropolarimetry with ESPaDOnS

The instrument includes an achromatic polarimeter, featuring rotatable retarders and installed at the Cassegrain focus of the 3.6 m Canada-France-Hawaii Telescope atop Mauna Kea. It fibre feeds a bench-mounted high-resolution spectrograph, yielding full coverage of the 370–1,000 nm wavelength range in a single exposure[14]. Polarimetric exposures consist of sequences of four subexposures collected at different orientations of the polarimeter retarders, each subexposure including both orthogonal states of the selected polarization. This procedure allows the suppression of all spurious polarization signals at first order[11]. The full efficiency of the instrument (atmosphere, telescope and detector included) is about 12%. Wavelength calibrated unpolarized and polarized spectra corresponding to each observing sequence are extracted with the dedicated software package Libre-ESpRIT, following the principles of optimal extraction[29].

### Deriving mean line profiles with least-squares deconvolution (LSD)

LSD is a cross-correlation type technique that allows us to obtain average unpolarized and polarized line profiles with largely enhanced signal to noise ratio (S/N), using simultaneously thousands of spectral features forming roughly in the same disk region[11]. Line lists are derived from spectrum synthesis through model atmospheres[30]. For this study, we mainly used a line list corresponding to a G0 spectral type (effective temperature of 6,000 K) and a logarithmic surface gravity of 2, which match best the spectrum of FU Ori[19,22]. A partial G0 line list including only the weakest spectral features, as well as line lists corresponding to spectral types A5 (8,000 K) and A0 (10,000 K) are also used in the analysis. LSD profiles are found to capture well the average shape of individual lines (and in particular the conspicuous flat-bottom aspect of weak lines, see Supplementary Fig. 2). With the full G0 line list, we obtain polarized LSD profiles with noise levels of about 0.01% (relative to the unpolarized continuum level) per 5.4 km s$^{-1}$ velocity bin, corresponding to a multiplex gain in S/N of more than 25 with respect to a single average line analysis. Higher noise levels, by factors of ~2 and ~4 respectively, are obtained with line lists corresponding to spectral types A5 and A0, containing much fewer spectral lines.

### Modelling the disk profiles

The model we use is inspired by theoretical magnetic disk models[5] and assumes axisymmetry. Ranging from 0.03 to 0.47 AU in radius, our model disk has a surface temperature (varying as $r^{-3/4}$, $r$ being the radial distance from disk centre) of ~7,000 K at 0.05 AU. For a mass accretion rate[8,13] of $10^{-4}$ $M_\odot$ yr$^{-1}$, a turbulence parameter[5] $\alpha_m \approx 1$ and a relative disk thickness[5] $h/r \approx 0.1$, the number density at disk midplane (varying with $r^{-3/2}$) is ~$10^{17}$ cm$^{-3}$ at 0.05 AU. The corresponding equipartition field (which varies as $r^{-5/4}$) is 1.5 kG at 0.05 AU. The local disk brightness is assumed to scale as a blackbody. The local absorption profile is modelled as a gaussian with a full-width at half-maximum of 15 km s$^{-1}$ and a temperature-dependent depth, whose variations are assumed gaussian with a mean and dispersion set to 6,500 and 2,500 K, respectively. We finally assume that only a fraction $f$ of the disk plasma is magnetic, that the field in the magnetic plasma is proportional to its equipartition distribution through the disk

(with scaling factors $b_v$ and $b_a$ for the vertical and azimuthal field components, respectively) and that the magnetic plasma rotates at a fraction $k$ of the keplerian speed (the non-magnetic plasma being keplerian). With this simple four-parameter model, synthetic Stokes $I$ and $V$ profiles are obtained by integrating the Doppler-shifted spectral contributions from all disk regions (the angle of the disk rotation axis to the line-of-sight being set to 60°, as estimated from interferometry[13]).

**Estimating the disk field strength and orientation**

For an axisymmetric disk field configuration, the vertical and azimuthal field distributions produce Stokes $V$ signatures that are respectively antisymmetric and symmetric with respect to the mean disk radial velocity, while the radial field distribution produces a null Zeeman signature (and thus cannot be estimated). After constructing the horizontally mirrored version $V^*$ of the observed Zeeman signature $V$ (with respect to the disk radial velocity of 27 km s$^{-1}$; refs 19, 20), we compute the half-difference $V_a=(V-V^*)/2$ and half-sum $V_s=(V+V^*)/2$ from $V$ and $V^*$ and straightforwardly obtain the unique additive decomposition of $V=V_a+V_s$ into its antisymmetric and symmetric components $V_a$ and $V_s$ with respect to the disk radial velocity. The synthetic Zeeman signatures corresponding to the vertical and azimuthal model disk field distributions can then be matched to $V_a$ and $V_s$ directly (see Fig. 3).

1. Shakura, N. I. & Sunyaev, R. A. Black holes in binary systems. Observational appearance. *Astron. Astrophys.* **24**, 337–355 (1973).
2. Pringle, J. E. Accretion discs in astrophysics. *Annu. Rev. Astron. Astrophys.* **19**, 137–162 (1981).
3. Blandford, R. D. & Payne, D. G. Hydromagnetic flows from accretion discs and the production of radio jets. *Mon. Not. R. Astron. Soc.* **199**, 883–903 (1982).
4. Pelletier, G. & Pudritz, R. E. Hydromagnetic disk winds in young stellar objects and active galactic nuclei. *Astrophys. J.* **394**, 117–138 (1992).
5. Ferreira, J. Magnetically-driven jets from Keplerian accretion discs. *Astron. Astrophys.* **319**, 340–359 (1997).
6. Bridle, H. A. & Perley, A. R. Extragalactic radio jets. *Annu. Rev. Astron. Astrophys.* **22**, 319–358 (1984).
7. Ray, T. P., Mundt, R., Dyson, J. E., Falle, S. & Raga, A. C. HST observations of jets from young stars. *Astrophys. J.* **468**, L103–L106 (1996).
8. Hartmann, L. & Kenyon, S. J. The FU Orionis phenomenon. *Annu. Rev. Astron. Astrophys.* **34**, 207–240 (1996).
9. Hutawarakorn, B. & Cohen, R. J. Magnetic structure in the bipolar outflow source G 35.2–0.74N: MERLIN spectral line results. *Mon. Not. R. Astron. Soc.* **303**, 845–854 (1999).
10. Hutawarakorn, B. & Cohen, R. J. OH maser disc and magnetic field structure. *Mon. Not. R. Astron. Soc.* **357**, 338–344 (2005).

11. Donati, J.-F., Semel, M., Carter, B. D., Rees, D. E. & Cameron, A. C. Spectropolarimetric observations of active stars. *Mon. Not. R. Astron. Soc.* **291**, 658–682 (1997).

12. Donati, J.-F. *et al.* Dynamo processes and activity cycles of AB Dor, LQ Hya and HR 1099. *Mon. Not. R. Astron. Soc.* **345**, 1145–1186 (2003).

13. Malbet, F. *et al.* New insights on the AU-scale circumstellar structure of FU Orionis. *Astron. Astrophys.* **437**, 627–636 (2005).

14. Donati, J.-F. ESPaDOnS: An Echelle SpectroPolarimetric Device for the Observation of Stars at CFHT. *ASP Conf. Proc.* **307,** 41–50 (2003)</conf>

15. Reipurth, B. & Aspin, C. The FU Orionis binary system and the formation of close binaries. *Astrophys. J.* **608**, L68–L68 (2004).

16. Wang, H., Apai, D., Henning, T. & Pascucci, I. FU Orionis: a binary star? *Astrophys. J.* **601**, L83–L86 (2004).

17. Preston, G. A statistical investigation of the orientation of magnetic axes in the periodic magnetic variables. *Astrophys. J.* **150**, 547–550 (1967).

18. Johns-Krull, C. M. & Gafford, A. D. New tests of magnetospheric accretion in T Tauri stars. *Astrophys. J.* **573**, 685–698 (2002).

19. Herbig, G. H. Eruptive phenomena in early stellar evolution. *Astrophys. J.* **217**, 693–715 (1977).

20. Herbig, G. H., Petrov, P. P. & Duemmler, P. High-resolution spectroscopy of FU Orionis stars. *Astrophys. J.* **595**, 384–411 (2003).

21. Hartmann, L. & Calvet, N. Observational constraints on FU Ori winds. *Astron. J.* **109**, 1846–1855 (1995).

22. Kenyon, S. J., Hartmann, L. & Hewett, R. Accretion disk models for FU Orionis and V1057 Cygni: detailed comparisons between observations and theory. *Astrophys. J.* **325**, 231–251 (1988).

23. Ferreira, J. & Pelletier, G. Magnetized accretion-ejection structures III. Stellar and extragalactic jets as weakly dissipative disk outflows. *Astron. Astrophys.* **295**, 807–832 (1995).

24. Brandenburg, A., Nordlund, A., Stein, R. F. & Torkelsson, U. Dynamo-generated turbulence and large-scale magnetic fields in a Keplerian shear flow. *Astrophys. J.* **446**, 741–754 (1995).

25. von Rekowski, B., Brandenburg, A., Dobler, W. & Shukurov, A. Structured outflow from a dynamo active accretion disc. *Astron. Astrophys.* **398**, 825–844 (2003).

26. Dubrulle, B. *et al.* An hydrodynamic shear instability in stratified disks. *Astron. Astrophys.* **429**, 1–13 (2005).

27. Balbus, S. A. & Hawley, J. F. A powerful local shear instability in weakly magnetised discs. *Astrophys. J.* **376**, 214–233 (1991).

28. Keppens, R., Casse, F. & Goedbloed, J. P. Waves and instabilities in accretion disks: magnetohydrodynamic spectroscopic analysis. *Astrophys. J.* **569**, L121–L126 (2002).
29. Horne, K. An optimal extraction algorithm for CCD spectroscopy. *Publ. Astron. Soc. Pacif.* **98**, 609–617 (1986).
30. Kurucz, R. L. *Atlas 9 Stellar Atmospheres and Programs* [CD set], Smithsonian Astrophysical Observatory, Cambridge USA (1993)

**Supplementary Information** is linked to the online version of the paper at www.nature.com/nature.

**Acknowledgements** We thank the CFHT staff for help during the engineering runs of ESPaDOnS. We are grateful to F. Rincon and A.C. Cameron for comments on earlier versions of the manuscript. This paper is based on observations obtained at the Canada-France-Hawaii Telescope (CFHT) which is operated by the National Research Council of Canada, the Institut National des Science de l'Univers of the Centre National de la Recherche Scientifique of France, and the University of Hawaii.

**Author Information** Reprints and permissions information is available at npg.nature.com/reprintsandpermissions. The authors declare no competing financial interests. Correspondence and requests for materials should be addressed to J.-F.D. (donati@ast.obs-mip.fr).

**Figure 1 Magnetic field detection in the protostellar accretion disk FU Ori. a**, The circularly polarized (Stokes $V$) LSD profile of FU Ori (top curve, expanded by 100, normalised to the unpolarised continuum intensity $I_c$ and shifted by +1.05) shows a clear Zeeman signature in conjunction with the unpolarized (Stokes $I$) LSD profile (bottom curve). **b**, The LSD Stokes $V$ profiles obtained at the other two epochs (dashed and dotted lines) are compatible at noise level with the detected Zeeman signature (solid line), suggesting that the parent magnetic structure is grossly axisymmetric.

**Figure 2 Disk and wind contribution to the LSD profiles of FU Ori. a**, Whereas the unpolarized LSD profile derived with the partial G0 line list (weak spectral features only, red line, rescaled to the size of an average line of the full G0 list) is symmetric and centred on the disk radial velocity with no additional blue-shifted absorption from the wind[21], LSD profiles derived with the G0, A5 and A0 lists (black, blue and green lines) exhibit an increasingly larger contribution from the wind. **b**, No such changes with line list (at noise level, single pixel spikes reflecting the higher noise levels obtained with warmer line lists) are observed for the Zeeman signature, whose origin is thus attributable to the disk.

**Figure 3 Modelling the LSD profiles of FU Ori. a**, Matching the wings of the observed unpolarized disk profile (solid line: partial G0 list) requires a disk keplerian velocity of 65 km s$^{-1}$ at 0.05 AU (dash-dot line); matching the flat-bottom core requires that 20% of the disk plasma rotates 50–70% slower than the keplerian velocity (dashed line). **b**, The Zeeman signature (top curve) is split into its antisymmetric and symmetric components (middle and bottom curves, shifted by $-4\times10^{-4}$ and $-8\times10^{-4}$). Fitting the observations (thick solid line, full G0 list; thin solid line, partial G0 list; dashed line, model) requires the slowly rotating disk plasma to be threaded by a 1 kG vertical field plus a 0.5 kG azimuthal field (at 0.05 AU). All profiles are shown in the disk velocity rest frame.

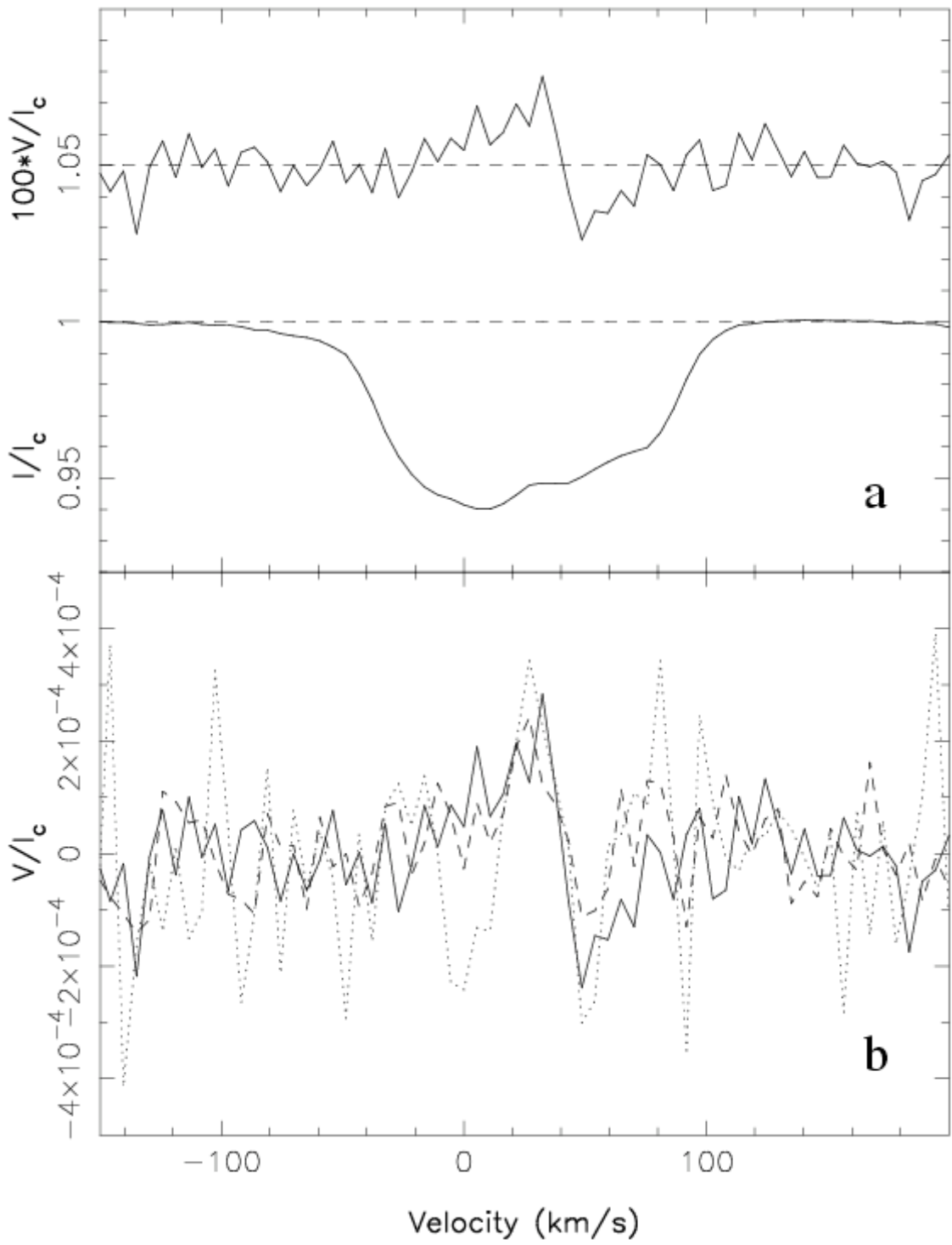

**Figure 1**

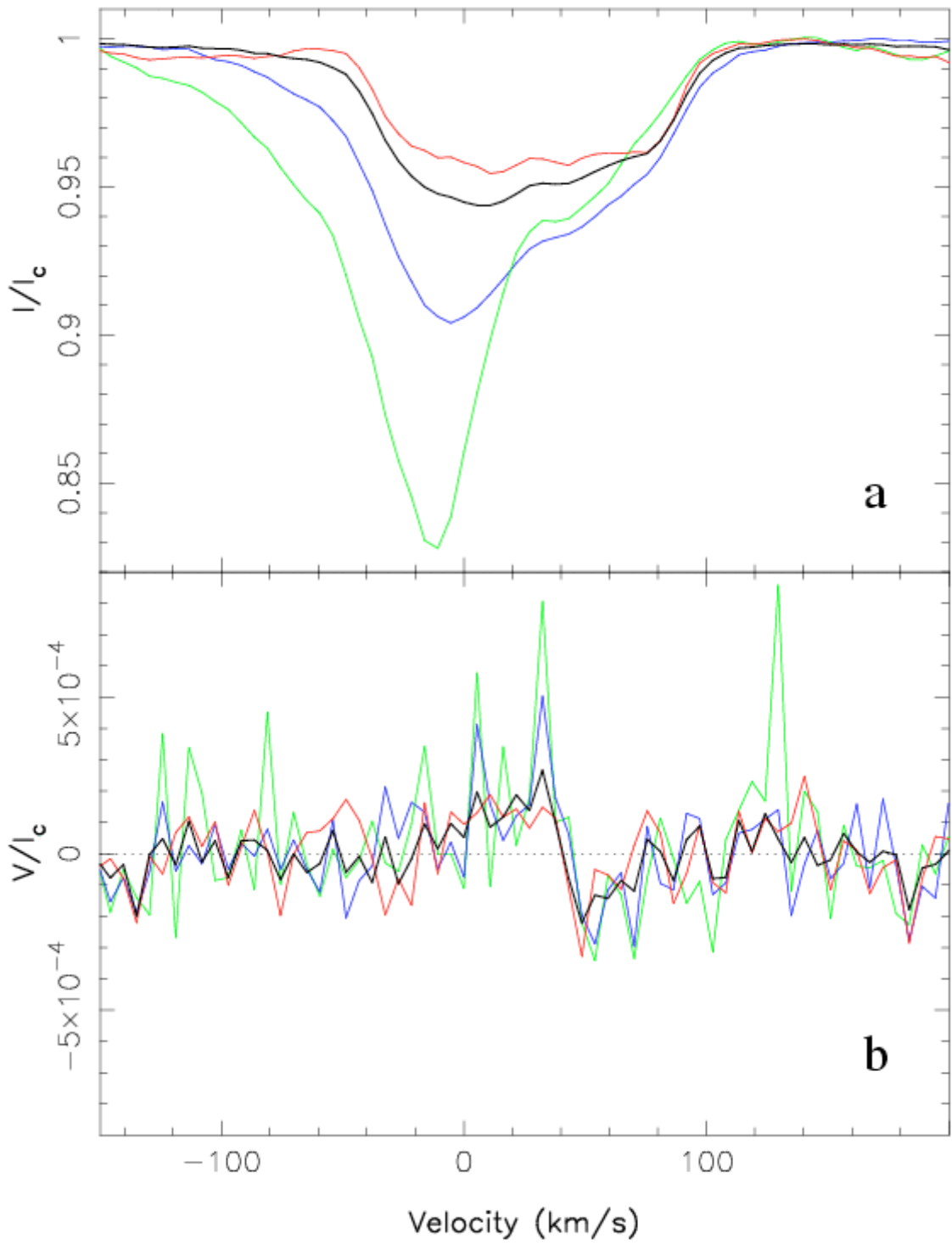

**Figure 2**

11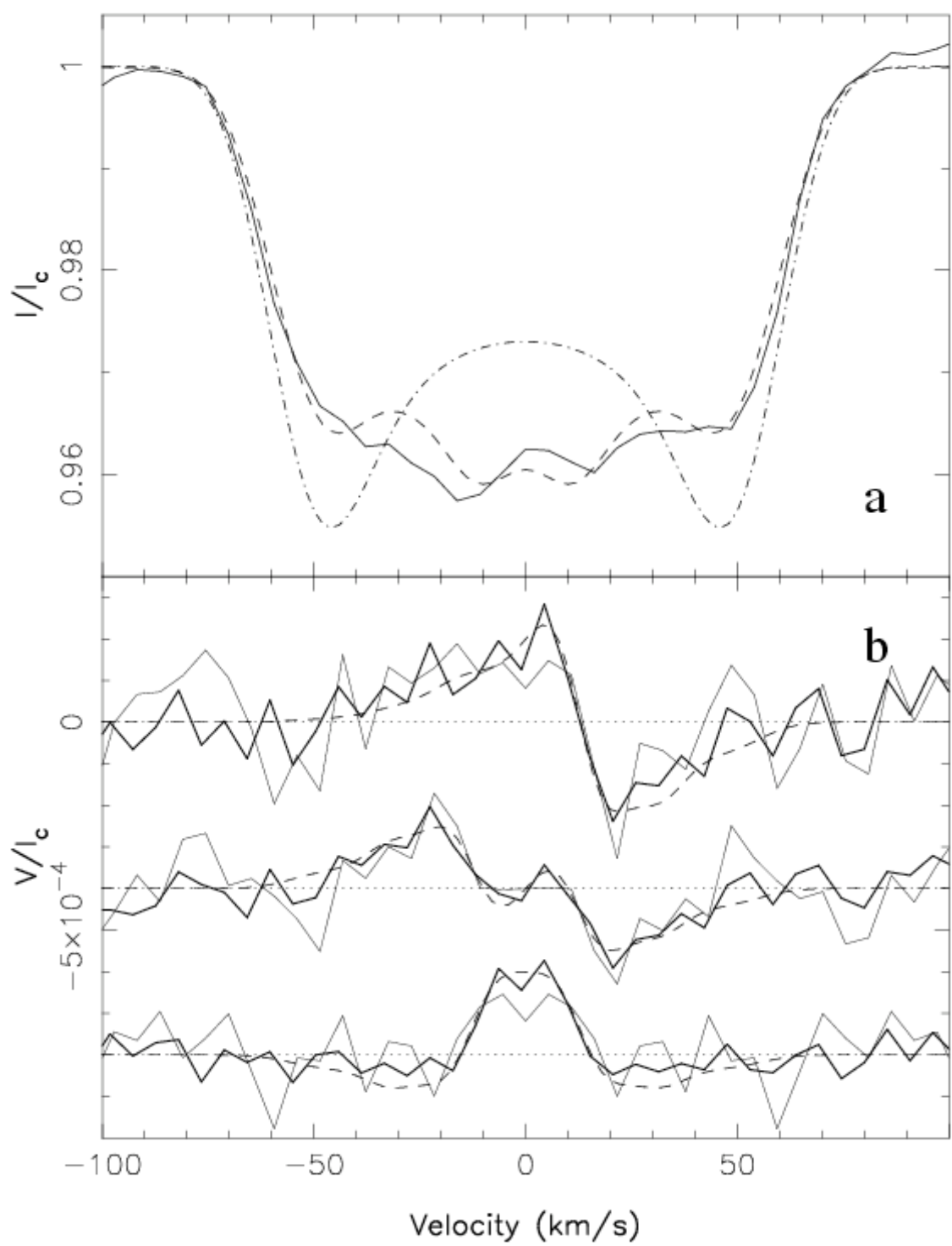

**Figure 3**